# An ECG-SoC with 535nW/Channel Lossless Data Compression for Wearable Sensors

C.J. Deepu, X. Zhang, W.-S. Liew, D.L.T. Wong, and Y. Lian
Department of Electrical & Computer Engineering, National University of Singapore
eleliany@nus.edu.sg

*Abstract*— This paper presents a low power ECG recording System-on-Chip (SoC) with on-chip low complexity lossless ECG compression for data reduction in wireless/ambulatory ECG sensor devices. The proposed algorithm uses a linear slope predictor to estimate the ECG samples, and uses a novel low complexity dynamic coding-packaging scheme to frame the resulting estimation error into fixed-length 16-bit format. The proposed technique achieves an average compression ratio of 2.25x on MIT/BIH ECG database. Implemented in 0.35μm process, the compressor uses 0.565K gates/channel occupying 0.4 mm$^2$ for 4-channel, and consumes 535nW/channel at 2.4V for ECG sampled at 512 Hz. Small size and ultra-low power consumption makes the proposed technique suitable for wearable ECG sensor application.

## I. INTRODUCTION

Cardiovascular disease (CVD) is the leading cause of death around the world and consumes significant amount of healthcare resources. An aging population, increasing life expectancies in developed countries are expected to aggravate this issue much further in the coming years. The effective way of managing CVD is to prevent it from happening by using low cost wearable wireless ECG (electrocardiogram) sensor. The main challenge in the development of a low cost wearable ECG sensor is the design of an ultra-low power ECG chip, which can acquire, process and wirelessly transmit ECG signal to a personal gateway, as shown in Fig. 1. A high level of integration, with inbuilt signal acquisition, data conversion [1], helps reduce the size and cost of such a sensor. In a wireless sensor, the single largest source of power consumption is the wireless transceiver. Due to round the clock operation, a large amount of ECG data has to be either locally stored on the device in a flash memory or transmitted wirelessly to a personal gateway, resulting in large memory and energy requirements at the sensor. In some cases an on-chip SRAM is used to facilitate the burst mode transmission resulting large chip area[2], which increases the cost of the device. Data compression offers several attractive features for such wearable sensors. It helps to reduce the size of on chip SRAM or Flash and minimize the power consumption of wireless transceiver. Although lossy compression techniques provides higher compression ratios[3], it may not be an ideal candidate for ECG signal due to the loss of information. Furthermore, lossy techniques are yet to be approved by medical regulatory bodies in many countries and can't be used in commercial devices due to liability concerns. So we prefer lossless compression for wearable ECG sensor. Most of the existing lossless ECG compression techniques are predominantly focused on achieving higher compression ratio (CR).

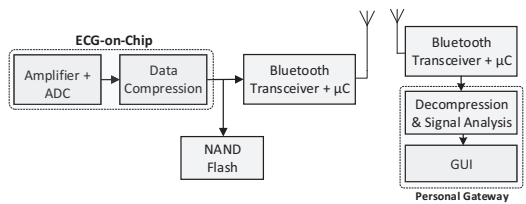

Fig.1 Wireless ECG Monitoring System

However in the context of wireless sensors and ambulatory devices, ultra-low power operation, low complexity in implementation are very critical. This is to make sure that the energy and memory savings from the compression is higher than what is consumed by the compressor itself.

In this paper, an ECG SoC featuring a low complexity lossless compression technique is presented. The proposed algorithm uses a slope predictor to estimate ECG samples, and uses a novel low complexity dynamic packaging scheme to frame the resulting estimation error into fixed-length format. The paper is organized as follows. In Section 2, the system architecture is presented. The compression scheme and performance evaluation are given in Section III. Section IV describes the hardware implementation. Measurement results are shown in Section V. Conclusions are drawn in Section VI.

## II. SYSTEM ARCHITECTURE OF ECG SOC CHIP

The system block diagram of the proposed ECG SoC is shown in Fig. 2. The frontend consists of 4 recording channels, a multiplexer (MUX) and a 12-bit successive approximation (SAR) ADC. The backend includes a lossless compression block, a real-time clock (RTC) module, and a SPI interface. To improve the ECG signal quality and reduce 50- or 60-Hz power-line noise, a driven-right-leg (DRL) is included. Also a low-power 32.768 kHz crystal oscillator driver and bandgap reference are implemented. The whole chip is designed to work under 2.4 ~ 3.0 V power supply.

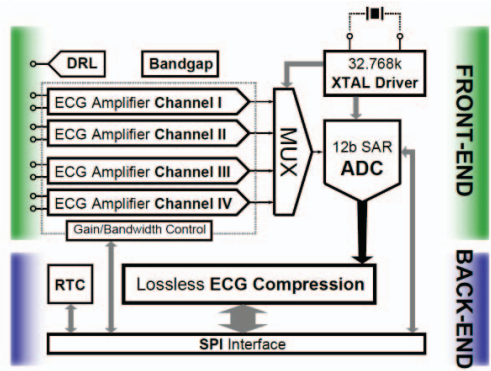

Fig. 2 System architecture of the ECG SoC

This work was supported in part by the National Research Foundation Competitive Research Programme under Grant NRF-CRP8-2011-01 and NUS Faculty Strategic Funding under Grant R-263-000-A02-731.

The architecture for the analog front-end amplifiers is shown in Fig. 3, which includes an instrumental amplifier (IA), a programmable gain amplifier (PGA), and a rail-to-rail output buffer (BUF). The capacitively coupled IA blocks the DC offsets and improves the signal dynamic range. The overall ECG gain is dynamically tunable through the PGA's switches T<1:0>. For antialiasing purpose, the low-pass cut-off frequency can be adjusted within 35 ~ 175 Hz, by changing the PGA's miller compensation capacitors. The unity-gain buffer improves the settling time for the MUX's output signal, reducing the residual errors [1]. Pseudo resistors used in IA and PGA provide to up GΩ resistance. This ensures the front-end high-pass frequency is less than 0.5 Hz, and the total harmonic distortion (THD) at full-scale output is less than 0.1%.

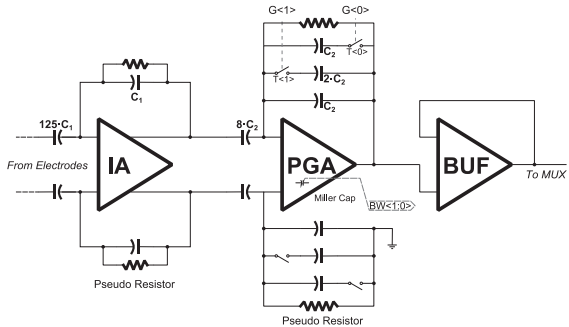

Fig.3 Architecture of a single ECG amplifier channel

### III. LOSSLESS DATA COMPRESSION SCHEME

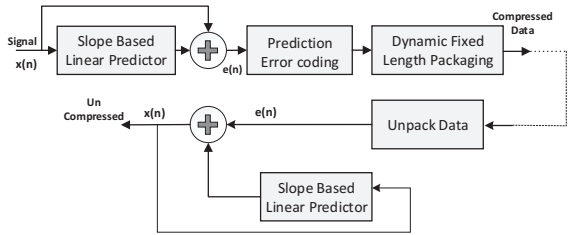

Fig.4 Lossless compression-decompression scheme.

The block diagram of the proposed compression – decompression scheme is illustrated in Fig. 4. We use a slope based predictor to estimate the value of the current ECG ($x(n)$) sample based on previous 2 samples ($x(n-1)$, $x(n-2)$), i.e. we compute the current sample value using the same signal slope obtained from previous 2 samples. This predicted value is subtracted from the actual value to obtain the prediction error of the current sample, The estimation for slope predictor and prediction error are given by (1), (2).

$$\hat{x}(n) = 2 * x(n-1) - x(n-2), \quad (1)$$
$$e(n) = x(n) - \hat{x}(n), \quad (2)$$

where $\hat{x}(n)$ is prediction estimate, $e(n)$ is the prediction error. The dynamic range of $e(n)$ is generally much smaller than that of the ECG signal as shown in Fig. 5. Noted that, for achieving lossless compression, it needs maximum ($L$+2) bits to fully represent $e(n)$, where $L$ is the bit-width of $x(n)$. With the proposed scheme, only prediction error needs to be transmitted instead of original ECG samples. At the receiver side, the reverse process is done to reconstruct the original data samples as shown in Fig. 4.

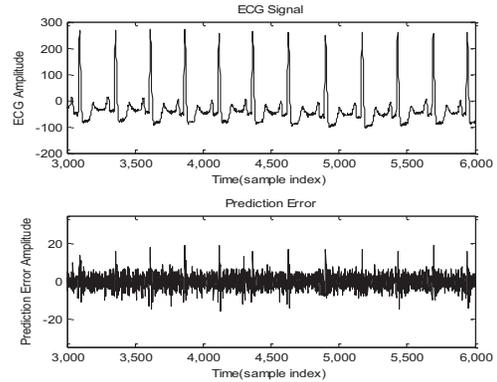

Fig.5 Prediction Error.

To reduce the bit-width of $e(n)$, a coding scheme can be used without incurring any data loss. *Variable length coding* schemes like Huffman and Arithmetic coding[4] are commonly used, which produces *prefix-free codes*[5] that can be packed closely. These approaches, although produce more or less optimal bit representations, the complexity of encoder/decoder is quite high[4][6]. For example, Huffman coding, associates the most frequently occurring symbols with short codewords and the less frequently occurring symbols with long codewords. This symbol-codeword association table has to be pre-constructed using a statistical dataset and the implementation of this table would require a large on chip memory[6], which ultimately would negate the effect of SRAM area savings from using compression. A suboptimal approach[5], selective Huffman coding, encodes only *m* frequently used symbols with Huffman codes and retains the remaining data un-encoded at the expense of a decrease in compression ratio. The hardware complexity of [5] is lower compared to the statistical approach. However it still need an *m* symbol lookup table at the encoder as well as decoder. In addition, these coding schemes produce, *variable length codes* at the output, that need further packaging to make it fit to fixed length packets, that can be stored in fixed word length SRAM/Flash or interfaced through a standard I/O like SPI. This re-packaging usually needs a complex hardware like the one proposed in [7].

We propose a simple coding-packaging scheme which combines coding and data packaging in one step. It has very low hardware complexity and achieves small area and low power while produces a fixed length 16-bit output. The flowchart of the scheme is presented in Fig. 6, where 2's complement representation ($e\_2c(n)$) of the error signal is used. As shown in Fig. 5, most error samples center around zero and hence can be represented by a few bits. Therefore we only select the necessary LSB's and remove any MSBs that doesn't carry information. However these data can't be packed closely, since it doesn't have the *prefix-free* nature of Huffman codes. For this, a data framing structure, as shown in Table I, with unique header for different frame types is used to pack the error samples of varying widths to a fixed length 16b output.

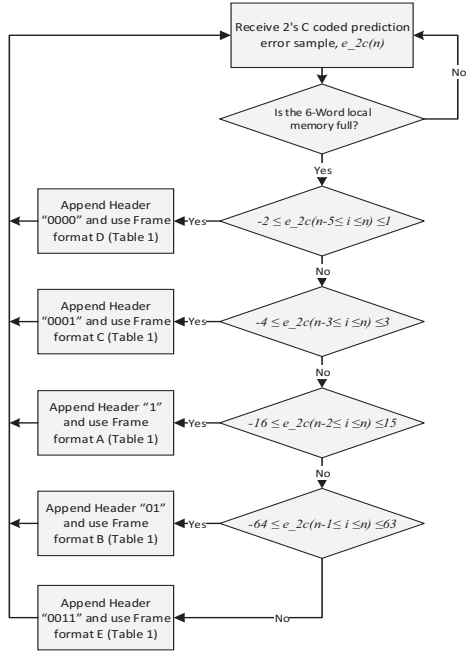

Fig 6. Coding-packaging Scheme Flowchart.

TABLE I
DATA PACKAGING SCHEME FOR 2'S CODED PREDICTION ERROR SYMBOLS.

| | | | | | | | | | |
|---|---|---|---|---|---|---|---|---|---|
| A | 1 | | 5/4 bits | | 5/4 bits | | | 5/4 bits | |
| B | 0 | 1 | | 7/6 bits | | | | 7/6 bits | |
| C | 0 | 0 | 0 | 1 | 3 bits | 3 bits | 3 bits | | 3 bits |
| D | 0 | 0 | 0 | 0 | 2 bits | 2 bit | 2 bits | 2 bits | 2 bits |
| E | 0 | 0 | 1 | 1 | | 12 bits | | | |

← 16 bits →

| Frame Type | Header | Data |

TABLE II
COMPRESSION PERFORMANCE OF THE PROPOSED ALGORITHM USING THE MIT/BIH DATABASE

| | Slope Predictor +Ideal Huffman | Slope Predictor +Selective Huffman | Slope Predictor +Proposed Scheme |
|---|---|---|---|
| Average CR | 2.66 | 2.16 | 2.25 |
| Maximum CR | 2.94 | 2.32 | 2.447 |

The dynamic data packaging scheme, in Fig. 6, uses a simple priority encoding technique to frame fixed length data from samples of multiple bit widths. As and when the error data is received, the algorithm checks whether the maximum amplitude of a signal group exceeds the value that particular frame format can accommodate from Table I. If not, the algorithm proceeds with the next best framing option.

The proposed algorithm is evaluated using the MIT/BIH recordings to obtain compression performance, as shown in Table II. The proposed coding scheme achieves 4% better performance than that of selective Huffman while produces fixed length coding. Its compression ratio is around 15% lower than ideal Huffman coding, but at significant less hardware cost as shown in next Section.

IV. HW ARCHITECTURE FOR PROPOSED COMPRESSOR

An overview of the hardware architecture for a 4-channel compressor is shown in Fig 7. It takes 12-bit, 512 S/s multiplexed 4-channel data from ADC output, and performances de-multiplexing before processing the data. The data from each channel is identified by a 2-bit channel select header appended to the ECG sample by the ADC. Each of the data stream is fed into a separate slope predictor for computing the prediction error. The individual predictor is activated with the channel select signal to reduce the effective data switching. The predictor data registers are set to zeros during initialization. These initial values are treated as the first 2 data samples when the system starts. Since the first 2 samples of the ECG stream are assumed to be zeros after initialization, there is no need to explicitly send the initial 2 samples for starting decompression. At the beginning, the minimum bit-width required for representing each prediction error in 2's complement format is computed. Since the packaging format shown in Table I only allows 5 types of data packets (2b, 3b, 5b, 7b and >8b), only these bit-widths are computed. The bit-width computation logic is shown in Fig. 8. Further the data is packaged into fixed 16 bit samples with bit packaging logic shown in Fig. 9. The error bit-width, along with corresponding error value, and actual samples are loaded serially into a 6-word register (Fig. 9). When the register is full, the packaging controller FSM, will pack the data from the register into a single 16-bit frame based on the inputs from Frame Enable logic. The order of priority is Frame type D, C, A, B (Table 1). The Frame enable logic for Frame type D is shown in Fig. 9.

In order to prevent total loss of data due to a few accidental packet loss, the proposed scheme allows to transmit full frames (Type E) at a predetermined interval. In our experiment, a 4s interval works well in office environment using Bluetooth. An SPI slave interface is used for interfacing the block with programmable MCU.

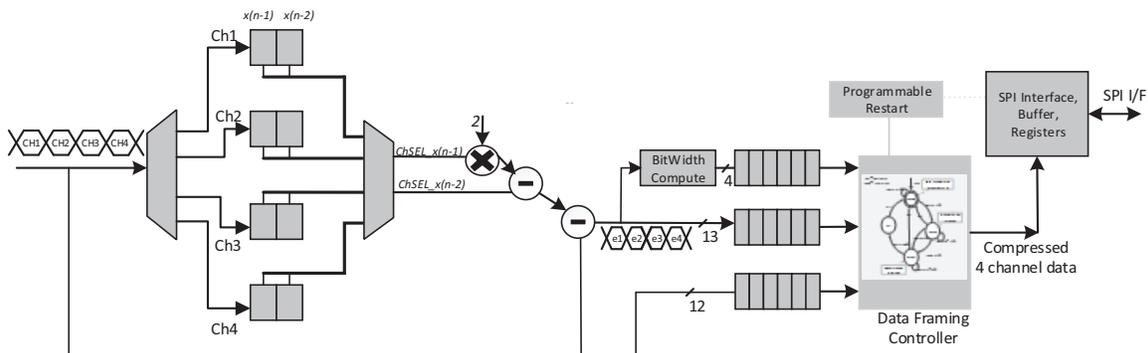

Fig 7. Hardware Architecture of the Lossless compression block for a 4 Channel ECG on Chip.

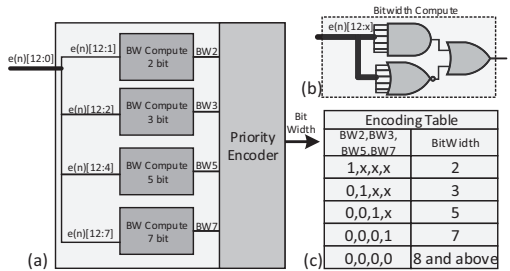

Fig 8. (a)Bit width block (b) Computation logic (c) Encoding Table

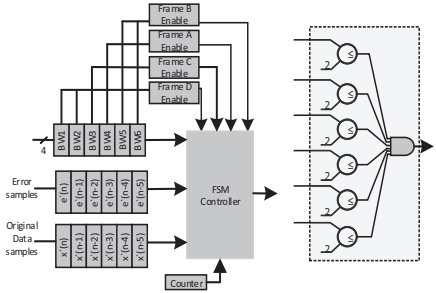

Fig 9. (a) Bit packaging block (b) Frame Enable for Type D frame

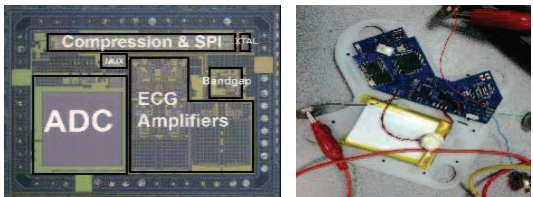

Fig.10 Chip micro photo and the prototype device.

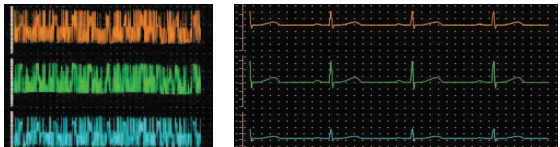

Fig.11 Before and after decompression.

## V. MEASUREMENT RESULTS AND COMPARISON

The design is implemented in a 0.35um process and the front-end measurement results are shown in Table III. The input referred noise from 0.05 Hz to 150 Hz is less than 1.5 µVrms. The high-pass corner is less than 0.01 Hz, and the full-scale 3V THD is only 0.08 %. The total front-end current, including all four channels, ADC, DRL, bandgap, and crystal oscillator circuit, is only 12.5 µA. The backend consumes 0.89 µA and has a core area of 0.2 × 2.0 mm$^2$. Fig. 10 shows the die photo and the evaluation prototype. The total chip area is 2.94 × 2.15 mm$^2$. Fig. 11 gives an example of the before- and after-compression ECG signal. No losses are observed from decompressed ECG signal.

Comparison of the proposed compressor with other recent published designs are given in Table IV. This design achieves the lowest power and complexity with a negligible reduction in compression ratio in chip measurements. [4] has better compression ratio, but consumes much higher power due to the implementation of Golomb coding scheme. [8] uses selective Huffman coding scheme and consumes much higher power. Note that [8] uses only 9 bit to represent prediction error, $e(n)$, for a 11 bit input data, when the data is left uncoded. For a fully lossless representation, $e(n)$ has to be represented by at least 13 bits. Else any prediction errors above 9 bits will be lost. Based on our study, we obtained 2.15x compression with selective Huffman coding scheme with a full bit-width representation for uncoded signals, as shown in Table II. Also the output from [8] has to be further packaged into fixed length for practical use. Overall the proposed design gives the lowest power and gate count for compression without compromising on data integrity.

TABLE III
CHIP PERFORMANCE SUMMARY

| | |
|---|---|
| Supply Voltage | 2.4~ 3.0 V |
| Technology | 0.35 µm CMOS |
| High-Pass Frequency | 0.0075 Hz |
| Low-Pass Frequency | 35 ~ 175 Hz |
| Pass-Band Gain | 47 ~ 66 dB |
| Input-Referred Noise (0.05 ~ 250 Hz) | 1.46 µVrms |
| Common-Mode Rejection Ratio (CMRR) | 65 dB |
| Power Supply Rejection Ratio (PSRR) | 76 dB |
| Total Harmonic Distortion (THD) | -64 dBFS |
| Sampling Frequency | 256/512 Hz |
| Total Front-End Current | 12.5 µA |
| Total Back-End Current | 0.89 µA |
| Effective Number of Bits (ENOB) | 9.3 |

TABLE IV
COMPARISON OF COMPRESSOR WITH OTHER DESIGNS

| | Ericson et al[4],2011 | S.L.Chen[8], 2013 | This design |
|---|---|---|---|
| Process | 65nm | 0.18µm | 0.35µm |
| Vdd | 1.0V | 1.8 V | 2.4~3V |
| CR | 2.38 | 2.43* | 2.25 |
| No of Channels | 1 | 1 | 4 |
| Total Gatecount | 53.9K | 3.57K | 2.26K |
| Gatecount/Channel | 53.9K | 3.57K | 0.56K |
| Total Power | 170µW | 36.4µW | 2.14µW |
| Power/Channel | 170µW | 36.4µW | 535nW |
| Area(mm$^2$) | 0.058 | 0.046 | 0.4 |

## VI. CONCLUSION

This paper presents a low power ECG SoC with lossless data compression for wearable devices. The compressor achieves an average CR of 2.25x. The design consumes 535nW/channel and has a core area of 0.4mm$^2$ in 0.35µm process. In comparison with other published methods, the proposed algorithm and hardware implementation demonstrates a low power consumption and is therefore suitable for wearable wireless devices.